\documentclass[prb,aps,twocolumn,groupedaddress,floats,final,superscriptaddress,nopacks]{revtex4}
\usepackage{graphicx}
\usepackage{subfigure}
\usepackage{amssymb}
\usepackage{latexsym}
\usepackage{epsfig}
\usepackage{psfrag}
\usepackage{dcolumn}
\usepackage{amsmath,amsfonts,amssymb,bm,ulem}
\usepackage{color}

\newcommand{\be}{\begin{equation}}
\newcommand{\ee}{\end{equation}}
\newcommand{\bea}{\begin{eqnarray}}
\newcommand{\eea}{\end{eqnarray}}

\newcommand{\s}{\sigma}

\newcommand{\la}{\langle}
\newcommand{\ra}{\rangle}
\newcommand{\rd}{\mbox{d}}
\newcommand{\ri}{\mbox{i}}

\begin{document}
\title{Resonant Inelastic X-Ray Scattering in Metals: A Diagrammatic Approach}

\author{A. M. Tsvelik}
\affiliation{Condensed Matter Physics and Materials Science Division, Brookhaven National Laboratory, Upton, NY 11973-5000, USA}
\author{R. M. Konik}
\affiliation{Condensed Matter Physics and Materials Science Division, Brookhaven National Laboratory, Upton, NY 11973-5000, USA}
\author{N. V. Prokof'ev}
\affiliation{Department of Physics, University of Massachusetts, Amherst, MA 01003, USA}
\author{I. S. Tupitsyn}
\affiliation{Department of Physics, University of Massachusetts, Amherst, MA 01003, USA}

\date{\today}

\begin{abstract}
We develop a formalism to study the Resonant Inelastic X-ray Scattering (RIXS) response in metals based on the diagrammatic expansion for its cross section. The standard approach to the solution of the RIXS problem relies on two key approximations: short-range potentials and non-interacting conduction electrons. However, these approximations are inaccurate for charged particles in metals, where the long-range Coulomb interaction and dynamic screening effects are very important. In this work we study how to extract important information about collective excitations in the Coulomb plasma, plasmons and electron-hole pairs, from RIXS data. We find that single- and multi-plasmon excitations can easily be distinguished by positions of the corresponding peaks, singularities, and their intensities. We also discuss the hybrid processes, where plasmon emission is accompanied by excitation of electron-hole pairs, and study how they manifest themselves.
\end{abstract}

\maketitle

\section{Introduction}

Resonant inelastic X-ray scattering (RIXS) technique holds tremendous promise for condensed matter physics \cite{dean,sawat2,chen,sawat4}. It possesses several unique features, and its ability to reach high energy and momentum transfer enables studies of a wide range of collective excitations (see, for example, Ref.~\cite{review}). However, the RIXS cross section is not proportional to the excitation spectral function, and thus extracting the necessary information from data requires proper understanding of the underlying processes. After an incident photon creates a deep core-hole, its strong potential disturbs the system and results in the emission of multiple excitations and, correspondingly, leads to various nonlinear effects. Therefore, to distill properties of excitations one needs to know how to separate single emission processes from the rest.

There exists a number of theoretical schemes to study the RIXS response \cite{rehr,demler1,demler2,NatureTD}. In this work we take a different track and develop a field-theoretical approach based on a diagrammatic expansion of the RIXS cross section (see also Refs.~\cite{Nomura,PlatIsa,phononsTD}). The advantage of this framework is in its universality---it allows one to address virtually any question about the system's behavior. Since RIXS is a second-order process (absorption followed by emission) we need to deal with the four-point time ordered correlation function $\chi_{R}(\omega_{i}, {\bf q}_i, \omega_{f}, {\bf q}_f)$ for the dipole operators (see, for instance, Ref.~\cite{demler1}) where $i$ and $f$ stand for the initial (incoming) and final (outgoing) photons. Such a formulation is similar to the one used to study the Raman scattering process \cite{shvaika}.


The overwhelming majority of theoretical efforts on the RIXS problem are based on two approximations:
(i) a static short-range/contact core-hole potential, and (ii) a non-interacting Fermi-sea.
However, both approximations oversimplify the nature of long-range Coulomb interactions and dynamic screening effects in metals. In the absence of dynamic screening, they lead to divergences that can be eliminated only by considering Fermi sea electrons as interacting via Coulomb forces as well.
As far as we know, the RIXS problem in metals has never been systematically addressed
beyond the formulation based on the above two approximations with the notable exception of Ref.~\cite{NatureTD} which incorporated Coulomb interactions into the dielectric function to study single-excitation emission process in layered copper-oxide systems.

The main goal of this work is to develop a more accurate understanding of various processes
based on the Coulomb interactions, including emission of multiple excitations, and establish
the framework for high-order diagrammatic expansion to the RIXS response (see, e.g. \cite{OC2019}).

There are two RIXS scenarios. In the first one, termed {\it indirect} RIXS, a deep-core $s$-electron is excited to a high energy, potentially mobile, $p$-state. The localized $s$-hole possesses a strong potential generating low energy collective excitations in the $d$-band (see Ref.~\cite{rehr}). In this case we have an $s-p$ dipole, emitting $d$-excitations during its life time, $\Gamma^{-1}$. The $p$-electron eventually repopulates the $s$-hole through a photon emission, leaving the $d$-excitations behind.
In the {\it direct} RIXS scenario, an electron from the $s$-band is excited into the $d$-band. Together $d$-electron and $s$-hole create collective excitations. During the hole's life time the excited $d$-electron moves away from the hole, and the photon is emitted when an electron from the occupied states recombines with the hole. In what follows we focus on the indirect RIXS process (direct RIXS is briefly commented in Conclusions). 

We work at $T=0$ where energy transfer to the system is always positive. Typically, the core-hole life time is very short---of the order of a few femtoseconds \cite{review}. This allows one to limit the diagrammatic expansion for the RIXS cross section to just a few collective excitations. Below we take advantage of short hole's life time (SHLT) and describe the dynamic screening within the Random Phase Approximation (RPA). For perturbative values of the Coulomb parameter $r_s$, this approach becomes exact. We concentrate on studying charge fluctuations (plasmons and particle-hole excitations) and discuss how signals from these collective excitations can be extracted from RIXS measurements by quantifying the contribution from the continuum of multiple excitations.

\section{The Anderson model for indirect RIXS}

Our diagrammatic expansion for the RIXS cross section follows the standard scheme (see \cite{shvaika}) which can be illustrated by considering the Anderson model for core-holes. To compute the cross section one introduces two species of holes (labeled by $a =1,2$) localized at different space points (``sites") at distance ${\bf R}_{12}$ from each other, and two species of $p$-electrons. Then the Anderson model can be formulated as follows:
\bea
H &=& H_s + H_d + H_p + H_{dd}  + H_{sp} + H_{sd} + H_{pd} ;   \\
H_s &=& \sum_{\s, a=1,2} \epsilon_s \; s^{\dagger}_{\s,a} \; s^{\;}_{\s,a} + H_{s,\Gamma}; \nonumber\\
H_p &=& \! \sum_{k, \s, a} \epsilon_p({\bf k}) \; p^{\dagger}_{{\bf k},\s , a} \; p^{\;}_{{\bf k},\s , a}; \;
H_d =   \! \sum_{k \s} \epsilon_d({\bf k}) \; d^{\dagger}_{{\bf k},\s} \; d^{\;}_{{\bf k},\s};  \nonumber \\
H_{sd} &=& -\int d{\bf r} \, n_d({\bf r})\times \Big[\frac{e^2}{|{\bf R}_{1}- {\bf r}|} \, n_{s,1}
+ \frac{e^2}{|{\bf R}_{2} - {\bf r}|} \, n_{s,2}  \Big],
\nonumber
\label{anderson}
\eea
where $s_{\s, a}, \; p_{\s , a}, \; d_{\s}$ are the annihilation operators for the $s$-core-, $p$-, and $d$-electrons (correspondingly, $s_{\s, a}$ creates the $s$-hole), $\sigma=\pm$ is the spin index, $\epsilon_s$ (here $\epsilon_s$ is $k$-independent),  $\epsilon_p({\bf k})$,  $\epsilon_d({\bf k}) $ are the corresponding dispersion relations, and $n_{d}$, $n_{s}$ are number densities. $H_{s,\Gamma}$ defines the s-hole with a finite lifetime $\Gamma^{-1}$. The interaction Hamiltonians $H_{sd}$, $H_{pd}$, and $H_{dd}$ have similar structure based on the Coulomb potential, $V_{{\bf r}} = e^2/r$ or  $V_{{\bf Q}} = 4\pi e^2/Q^2$ (for brevity, we present explicitly only $H_{sd}$). Formally, in the orbital representation, these interactions are different, but this difference is insignificant for the purposes of our work.

\begin{figure}[htp]
\vspace{-2mm}
\centerline{\includegraphics[angle = 0,width=0.9\columnwidth]{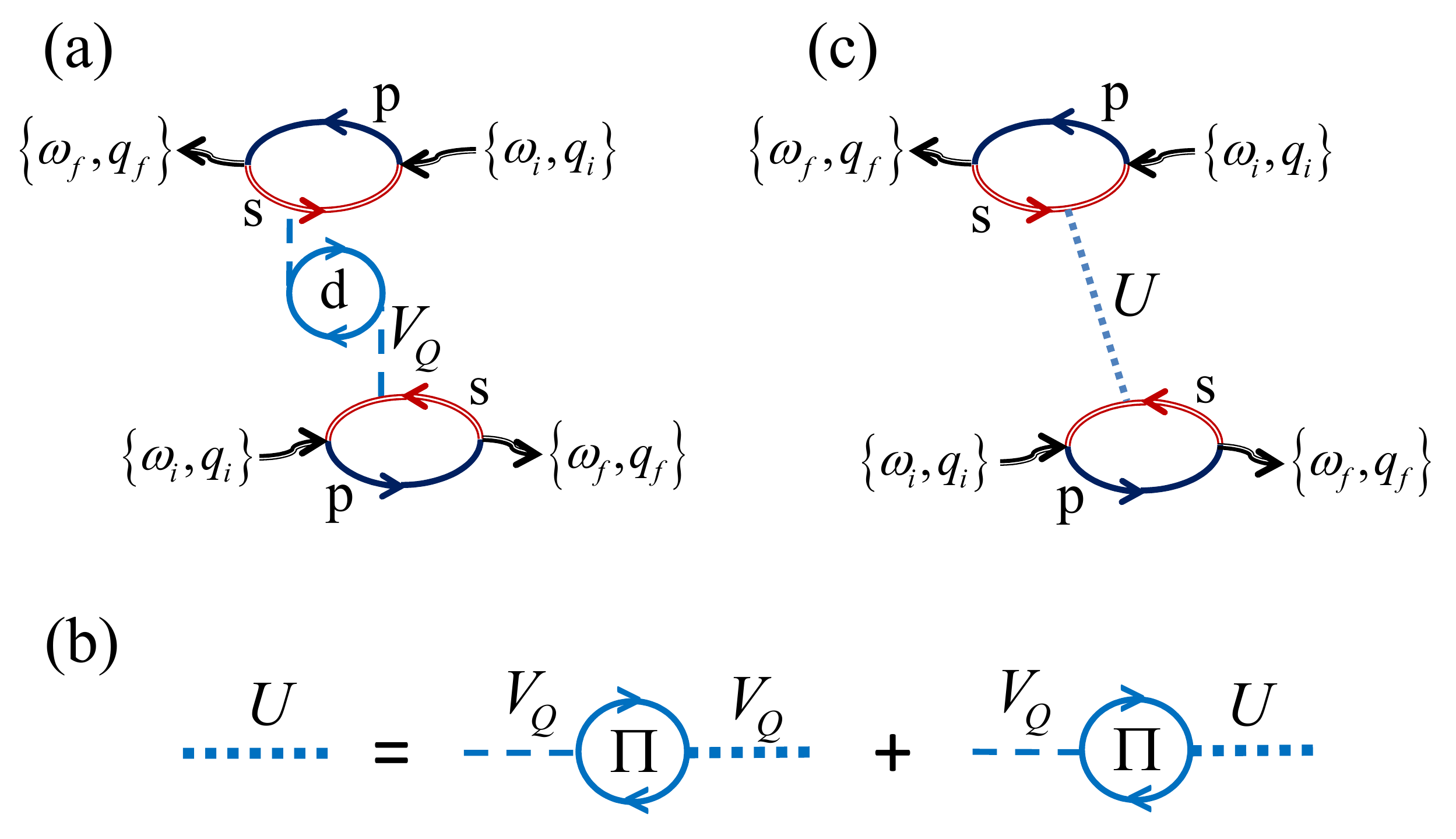}}
\vspace{-3mm}
\caption{ (color online). (a) The lowest-order diagram for the indirect RIXS cross-section where the $s$-hole excites charge density fluctuations in the $d$-shell (the internal fermionic bubble) via the Coulomb potential $V_Q$ (dashed lines). $(\omega_i,{\bf q}_i)/(\omega_f, {\bf q}_f)$ are the frequencies and momenta of incoming/outgoing photons. Energy and momentum transfer to the system are defined by $\Omega=\omega_i - \omega_f$ and ${\bf Q} ={\bf q}_i - {\bf q}_f$, respectively.
(b) The Dyson equation for the screened dynamic interaction $U$ obtained by summing up bubble diagrams based on $V_Q$ and the polarization function, $\Pi$, of $d$-electrons.
(c) The lowest-order diagram for the indirect RIXS cross-section in terms of $U$.  }
\label{Fig1}
\end{figure}

By integrating $d$-electrons out within the RPA, we arrive at the model where $s$-holes/$p$-electrons are coupled by the action
\bea
&&S = S_0 + \int \rd \tau_1\rd \tau_2 d{\bf r}_1 d{\bf r}_2
\rho({\bf r}_1,\tau_1)U( {\bf r}_{12}, \tau_{12}) \rho({\bf r}_2, \tau_2), \nonumber \\
&&\rho({\bf r}) = \sum_{a} \left[ \delta({\bf r} - {\bf R}_a) n_{s,a} + n_{p,a} ({\bf r}) \right],
\;\;\;\;\;\;\;\;\;\;\;\;\;\;\;\;\;\;\;\;\;\;
\label{anderson2}
\eea
where $S_0$ is the bare action for $s$-holes/$p$-electrons and $U$ is the dynamic part of the screened Coulomb potential (see Fig.~\ref{Fig1}(b)). In this formulation, the correlation function, $\chi_{R}$, responsible for the RIXS cross section, can be written as
\bea
&& \chi_{R}({\bf R}_{12}; t_1,t_2,t) = \label{corr} \\
&& \la D_1(-t_1/2) \; D^+_1(t_1/2) \; D_2(t_2/2 + t) \; D^{+}_2(-t_2/2 + t) \ra, \nonumber
\eea
where $D^+_a =p^{+}_{\s, a} s_{\s, a}$ is the dipole creation operator on site ${\bf R}_a$. The RIXS cross section is extracted from the imaginary part of the analytically continued Fourier transform of this correlation function in direct analogy to the Raman scattering response \cite{shvaika}:
\bea
&& \chi_{R}(\omega_{i}, {\bf q}_i, \omega_{f}, {\bf q}_f) = \frac{1}{2\pi i} \,\lim_{\delta_{1} > \delta_{2} \rightarrow 0}
\label{chi} \\
&& \Big( \tilde{\chi}(X_{1,i}; X_{1,f}; X_{2,f}; X_{2,i}) -\tilde{\chi}(X^{'}_{1,i}; X^{'}_{1,f}; X^{'}_{2,f}; X^{'}_{2,i}) \Big), \nonumber \\
&&  X_{1,i}     = \{ -\omega_{i}-i\delta_{1},-{\bf q}_i \};  \;\;
    X^{'}_{1,i} = \{ -\omega_{i}-i\delta_{2},-{\bf q}_i \}; \nonumber \\
&&  X_{1,f}     = \{  \omega_{f}+i\delta_{2}, {\bf q}_f \};  \;\;
    X^{'}_{1,f} = \{  \omega_{f}+i\delta_{1}, {\bf q}_f  \};  \nonumber \\
&&  X_{2,f}     = \{ -\omega_{f}+i\delta_{2},-{\bf q}_f \};  \;\;
    X^{'}_{2,f} = \{ -\omega_{f}+i\delta_{1},-{\bf q}_f  \}; \nonumber \\
&&  X_{2,i}     = \{  \omega_{i}-i\delta_{1},{\bf q}_i \};   \;\;
    X^{'}_{2,i} = \{  \omega_{i}-i\delta_{2},{\bf q}_i \}. \nonumber
\eea
Self-explanatory notations for all variables are given in Figs.~\ref{Fig1} and \ref{Fig2}.
In what follows, we expand $\chi_{R}$ into a diagrammatic series. The fully dressed Green's functions for both the $s$-hole and the $p$-electron (to account for interactions with $d$-electrons) can be obtained within the Diagrammatic Monte Carlo technique for polarons \cite{OC2019,Polaron1998}. However, under the SHLT assumption, they can be approximated by their non-interacting expressions.

\section{Diagrammatic representation for the cross section $\chi_{R}$}

If we focus on charge excitations for indirect RIXS, the relevant diagrammatic expansion is in the number of non-local interaction lines (\ref{anderson2}). The lowest order Feynman diagram for $\chi_{R}$ is shown in Fig.~\ref{Fig1}(a). The same diagram, but with a contact interaction $V$ instead of the Coulomb potential $V_Q$, was considered in \cite{Nomura} for the case of quasi-$1$d Mott insulators.
In contrast, Eq.~(\ref{anderson2}) is written in terms of the dynamic interaction $U$ based on the geometric series of bubble diagrams, as shown in  Fig.~\ref{Fig1}(b). In the RPA, the polarization function $\Pi$ is obtained from the product of two bare Green's functions for $d$-electrons. While certainly insufficiently accurate for, say, transition metals, the RPA (exact in the limit of small $r_s$) does capture all
qualitative features of the screening phenomenon. Thus, for Coulomb systems, the expansion order is defined by the number of the $U$ lines.

\begin{figure}[htp]
\vspace{-2mm}
\centerline{\includegraphics[angle = 0,width=0.9\columnwidth]{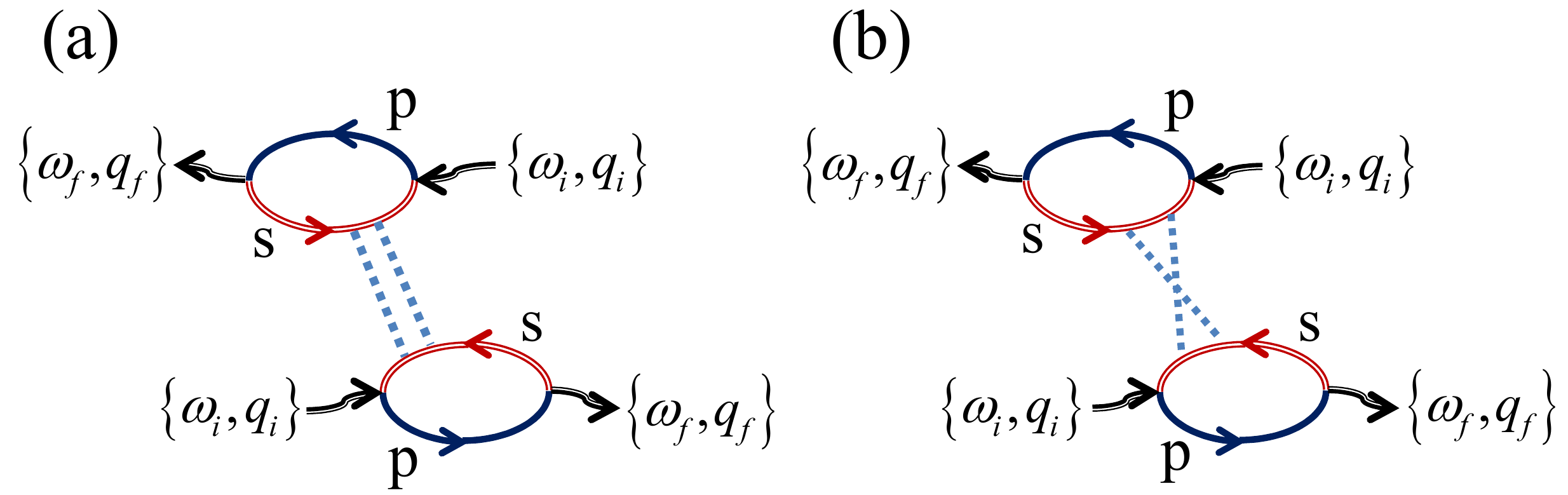}}
\vspace{-2mm}
\caption{(color online). Second-order diagrams for the RIXS cross section
with two different topologies (a) and (b).
}
\label{Fig2}
\end{figure}

In the SHLT limit it is enough to keep only few terms in the expansion.  Two second-order diagrams are presented in Fig.~\ref{Fig2}. Calculation of the vertex functions, $ \gamma_v$, to leading order in $1/\Gamma$ is presented in the Appendix. The contribution from the $s-p$ loops does not depend on energy transfer $\Omega$, but its momentum dependence is important. Since the $d$-electrons interact with a neutral $s-p$ excitation, all vertex functions vanish at zero momentum transfer. In particular, the vertex function squared for one $U(Q)$ line, Fig.~\ref{Fig1}c, contains a factor of $Q^4$, while for the sum of two diagrams with $U(q_1)$ and $U(q_2)$ lines shown in Fig.~\ref{Fig2}, we have $ |\gamma_v^{(2)}|^2 \propto ({\bf q}_1{\bf q_2})^2$. This difference leads to the suppression of contribution from the coherent plasmon at small momenta. The vertex function for the $n$-th order diagram is also proportional to $ \Gamma^{-(n+1)}$.

The general structure of diagrams implies that emission of varying number of gapped excitations leads to different thresholds. This, in principle, allows one to distinguish the processes accompanied by additional excitations, provided that the spectral gaps are not too small compared to the bandwidths of these excitations. Given that contributions from higher-order diagrams are suppressed by a factor $\propto \Gamma^{-2(n-1)}$, the most important corrections are determined by the second-order diagrams, see Fig.~\ref{Fig2}. For this reason the main focus of this work is on the second-order processes.
Note that the second-order contribution may dominate in the final answer if the first-order process
has zero intensity in some frequency range, or at small momentum transfer.

\section{Plasmon and electron-hole excitations}

In this Section we explore how important information about charged excitations in metals
can be extracted from the RIXS data by comparing first-order and second-order processes.
In the SHLT limit, the imaginary part of the correlation function responsible for the RIXS signal
originates from the $U$ functions, not from the vertex functions.
Then, the first-order contribution can be written as
\bea
\chi^{(1)}_{R} =  |\gamma_v^{(1)}(\omega_{i},\omega_f;\; {\bf Q}) |^2 D^{(1)}(\Omega ,Q),
\eea
where $D^{(1)}=U''$ is the imaginary part of the screened potential
\be
U(\Omega, Q) = \frac{4\pi e^2}{Q^2 - 4\pi e^2 \Pi(\Omega,Q)} -V_Q \,.
\label{FullU}
\ee

\begin{figure}[tbh]
\centerline{\includegraphics[angle = 0,width=0.8\columnwidth]{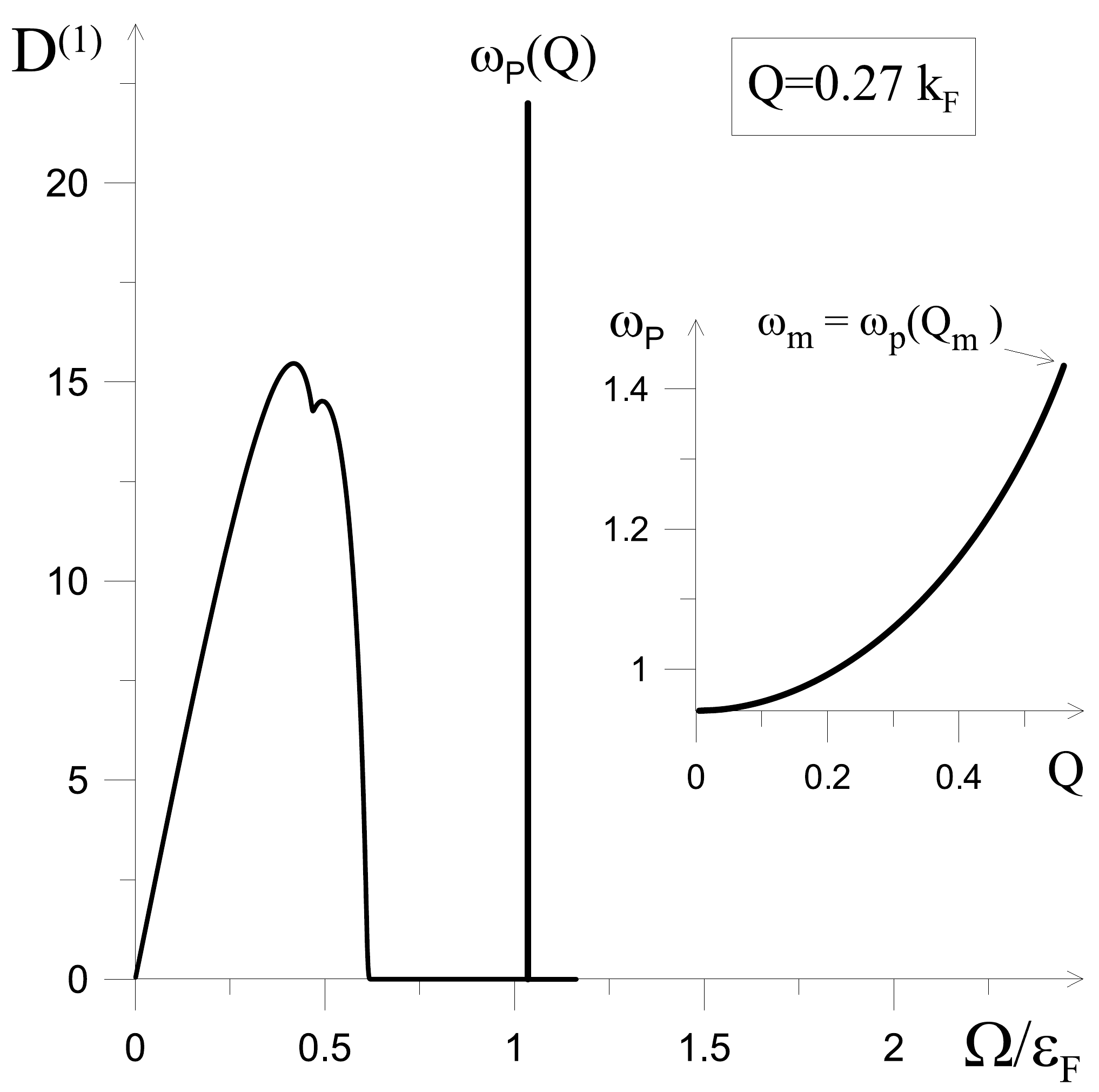}}
\caption{First-order, see Fig.~\ref{Fig1}(c), contribution to the RIXS intensity
(without the $|\gamma^{(1)}_v|^2$ factor) at $r_s=1$. The plasmon dispersion relation is shown in the inset.}
\label{Fig3}
\end{figure}

At zero temperature the energy transfer to the system is non-negative, and in what follows
we will implicitly assume that $\Omega \ge 0$. By approximating the $d$-band dispersion relation with the spherically symmetric expression $\epsilon_d = k^2/2m$, we easily obtain the result for $D^{(1)}(\Omega, Q)$ since $\Pi(\Omega,Q)$ in this situation reduces to the Lindhard function \cite{Lindhard} (a typical curve is shown in Fig.~\ref{Fig3}). We work with units
such that the Fermi momentum, $k_F=1$, and Fermi energy, $\epsilon_F=1$.

The main features of the $D^{(1)}$ spectrum are the sharp plasmon peak and the electron-hole continuum. Since plasmons are gapped excitations (their dispersion relation, $\omega_p(Q)$, is shown in the inset of Fig.~\ref{Fig3}), their contribution to intensity is separated from the low energy particle-hole continuum at small enough momenta, see Fig.~\ref{Fig3}. This separation becomes less pronounced at large momentum transfer and ultimately the plasmon peak merges with the continuum at
\be
\Omega_m=\omega_p(Q_m)=v_FQ_m+\frac{Q_m^2}{2m}.
\label{WM}
\ee

To recover the same basic properties we radically simplify the Lindhard function while preserving
exactly the key features of the important $Q \ll \kappa$ limit in terms of the Fermi liquid parameters.
The corresponding approximation combines the plasmon pole approximation \cite{Lundquist67,HedLund1969}
with Landau damping:
\be
\Pi \approx - \rho_F  + \frac{\rho_F \Omega}{2 v_F Q} \left[
\ln \left\vert \frac{\Omega + v_F Q}{\Omega - v_F Q} \right\vert
- \ri \pi \theta (v_F Q -\Omega) \right].
\label{PIRPA}
\ee
Furthermore, in the same limit, the imaginary part of $U(\Omega,Q)$ can be separated
into two distinct contributions, $U'' = D^{(1)}_{p-h} + D^{(1)}_{pl}$, associated with
excitation of low-energy particle-hole pairs and gapped plasmon modes, respectively:
\bea
D^{(1)}_{p-h} = \frac{4\pi e^2}{\kappa^2}\frac{\pi\Omega}{2v_FQ} \; \theta(v_FQ-\Omega), \label{one-ph} \\
D^{(1)}_{pl}  = \frac{4 \pi e^2}{Q^2}\frac{\pi}{2}\omega_{p}(Q) \; \delta(\Omega - \omega_{p}(Q)).
\label{one-pl}
\eea
Finally, by using
\be
\omega_{p}(Q) = \Omega_{pl} + \xi\,Q^2; \;\;\; \xi = \frac{3}{10} \frac{v_F^2}{\Omega_{pl}}
\label{Pl_disp}
\ee
in Eq.~(\ref{one-pl}), we correctly capture the plasmon dispersion at low momenta.
By developing this effective description we are now in position to address the problem of
emission of multiple excitations in order to see whether and how their contributions can be separated from the
first-order single-emission process.

\begin{figure}[htp]
\centerline{\includegraphics[angle = 0,width=0.8\columnwidth]{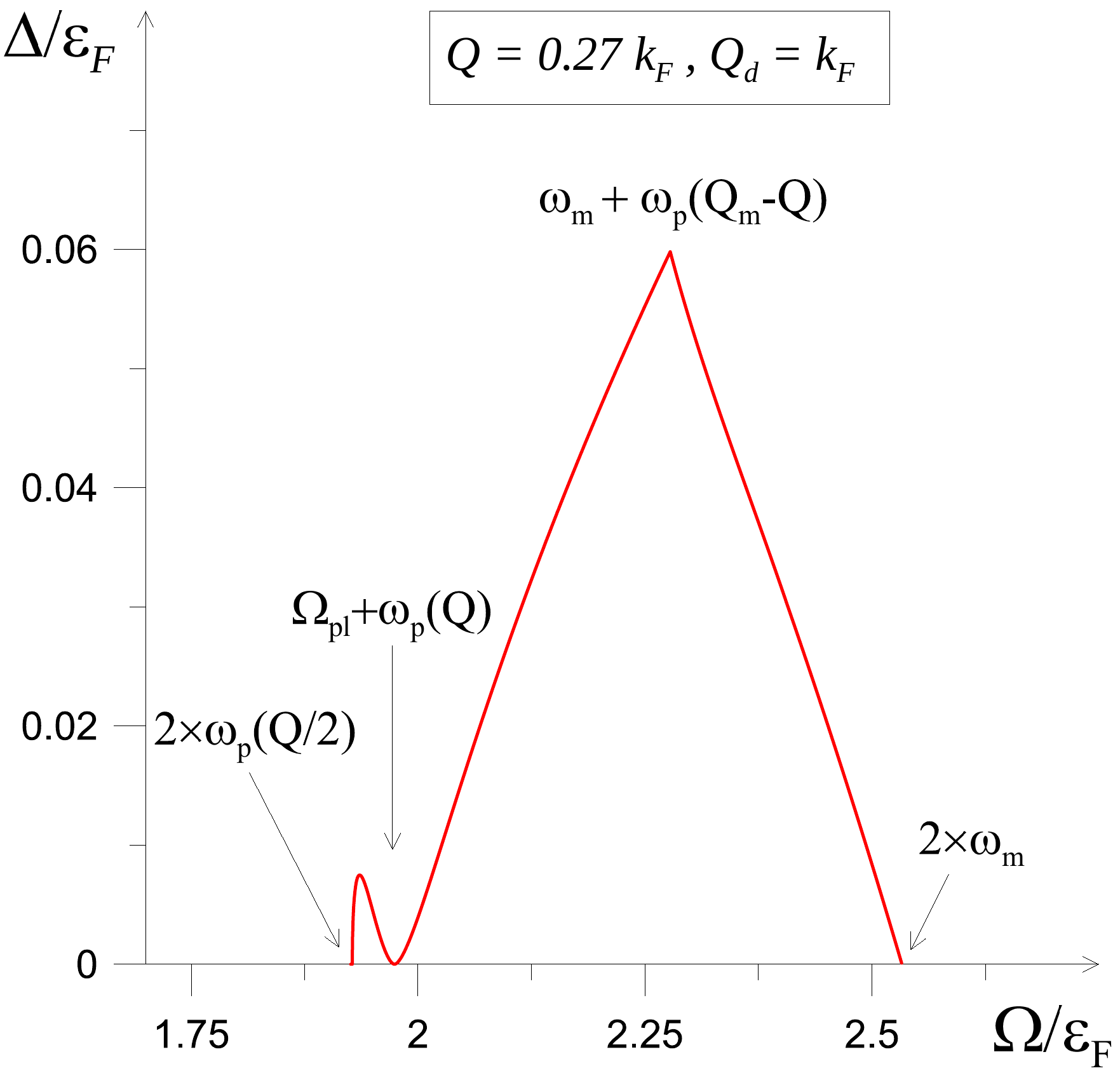}}
\caption{Second order contributions to the RIXS intensity: $\Delta$ as a function of frequency $\Omega$ at $r_s=1$. [For the simplified plasmon dispersion (\ref{Pl_disp}), $Q_m \approx 0.5055k_F $.] }
\label{Fig4}
\end{figure}


In the $\Omega_{pl}/\Gamma << 1$ limit, the two diagrams shown in Fig.~\ref{Fig2} give the same contribution. To account for the momentum dependence of the vertex function we write it as
$|\gamma^{(2)}_v({\bf q}_1,{\bf q}_2)|^2 = (\gamma^2/\Gamma^2 ) \; f({\bf q}_1,{\bf q}_2)$
with constant $\gamma^2$ and
\be
f({\bf q}_1,{\bf q}_2) = \frac{q^2_1 \; q^2_2 \; {\cal C}^2_{12}}{(q^2_1+Q_d^2)(q^2_2+Q_d^2)},
\label{FQ1Q2}
\ee
where ${\cal C}_{12}$ is the cosine of the angle between vectors ${\bf q}_1$ and ${\bf q}_2$,
$Q_d \sim k_F$ is some high-momentum cutoff, and proceed with evaluating the integral
\be
\int  \frac{d{\bf q} d\omega}{(2\pi)^{4}}   f({\bf q},{\bf Q}-{\bf q})  D^{(1)} (\omega, {\bf q})
                                            D^{(1)} (\Omega -\omega, {\bf Q}-{\bf q}).
\label{integral2}
\ee
For two plasmons, the final result reads (without $\gamma^2$):
\be
D^{(2)}_{pl} \propto  \frac{\Omega_{pl}^2}{\Gamma^2}\, \frac{\pi e^4}{Q} \Delta(\Omega,Q)
\label{two}
\ee
with $\Delta$-function 
\bea
\Delta &=& \frac{1}{\Omega_{pl}^2} \int_0^{Q_m} \frac{qdq}{q^2+Q_d^2}\,
\frac{\omega_{p}(q) \; (\Omega - \omega_{p}(q)) \; {\cal C}^2_{12}} {[\Omega-\Omega_{pl}-\omega_{p}(q) + \xi Q_d^2]}
\nonumber \\
&\times& \Theta \left(1 - \left \vert \frac{2\omega_p(q)+\omega_p(Q)-\Omega_{pl}-\Omega}{2 \xi Q q} \right \vert \right); \\
{\cal C}^2_{12} &=& \frac{(\Omega-\Omega_{pl}-\omega_p(Q))^2}
{4[\omega_p(q)-\Omega_{pl}][\Omega-\Omega_{pl}-\omega_{p}(q)]} \nonumber
\eea
featuring singularities in derivatives distinctly related to the plasmon spectrum, see Fig.\ref{Fig4}. For a dispersionless plasmon, $\Delta(\Omega)$ is proportional to $\delta(\Omega - 2 \Omega_{pl})$. Thus, the one- and two-plasmon processes can be distinguished by positions and shapes of the corresponding peaks and intensities: while the one-plasmon process results in the sharp peak at $\omega_{p}(Q)$, the two-plasmon curve is broad and is shifted outside of the plasmon dispersion relation. The ratio of intensities goes as
\be
\frac{ \gamma^2 D^{(2)}_{pl} }{ |\gamma^{(1)}(Q)|^2 D^{(1)}_{pl} }   \propto
\frac{ \Omega_{pl} e^2 k_F^3 }  { \Gamma^2 Q^2 }.
\label{ratio}
\ee
At low momenta the two-plasmon process will produce a stronger signal than the single-plasmon one.

\begin{figure}[htp]
\centerline{\includegraphics[angle = 0,width=0.8\columnwidth]{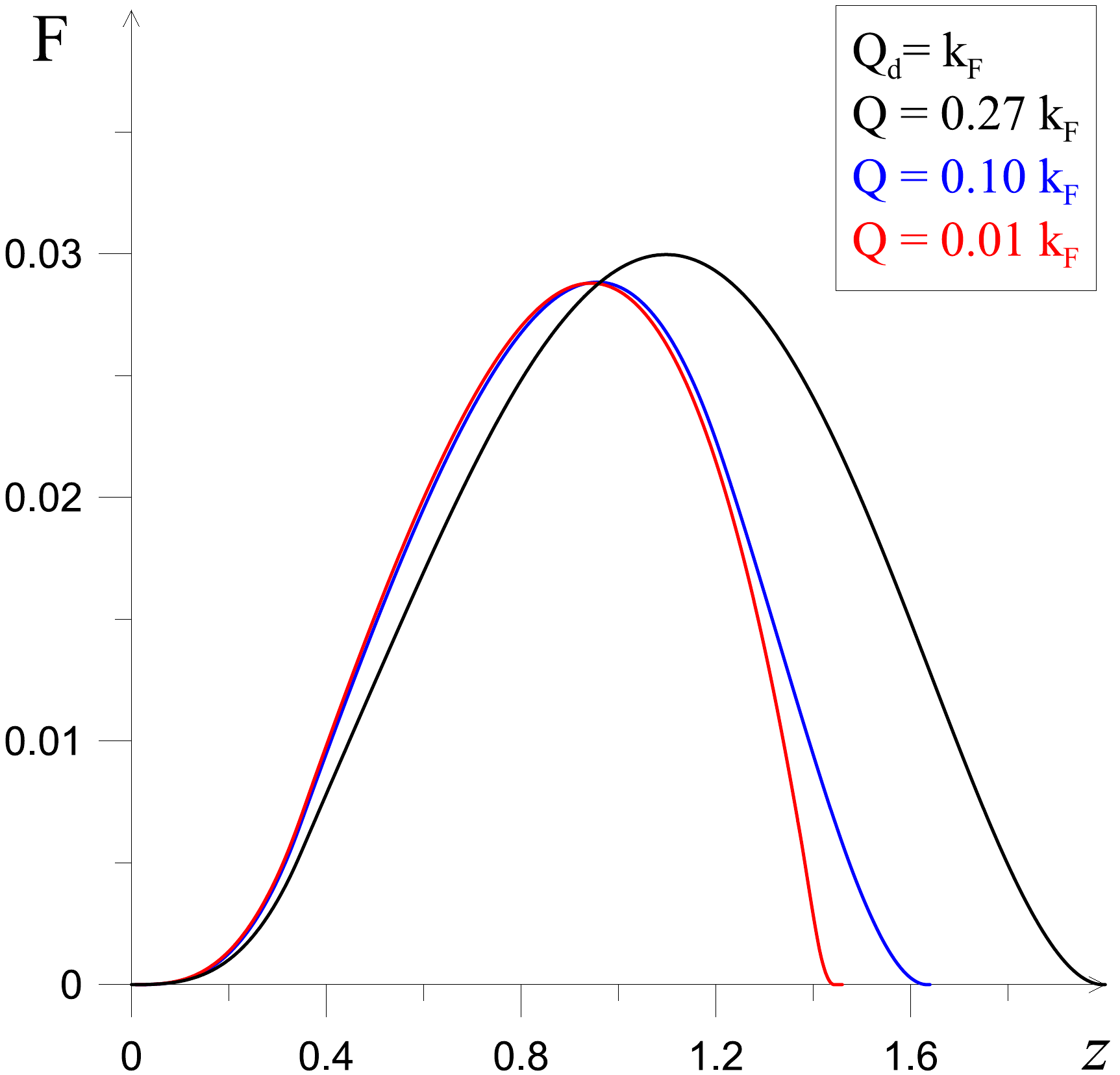}}
\caption{Second order contributions to the RIXS intensity: The hybrid amplitude $F$ as a function of $(\Omega-\Omega_{pl})/\Omega_{pl}$ for three values of external momentum transfer ($Q=0.27k_F$--black, $Q=0.1k_F$--blue, and $Q=0.01k_F$--red) at $r_s=1$.}
\label{Fig5}
\end{figure}

There is also a second-order hybrid process, when plasmon emission is accompanied
by excitation of the particle-hole pair. The corresponding spectrum overlaps with the single-plasmon peak:
\be
D^{(2)}_{pl,p-h} \propto \frac{\Omega_{pl}^2}{\Gamma^2} \; \frac{\pi^2 e^2}{4k_F^2} \; F(z,y),
\label{correction}
\ee
where $z=(\Omega-\Omega_{pl})/\Omega_{pl}$, $y=v_FQ/\Omega_{pl}$, and 
\bea
F(z,y) &=&
\int_{0}^{x_m} dx \frac{(1+x)(z-x)}{x (x + \xi Q_d^2/\Omega_{pl})} \label{correction2} \\
&\times& [Y(x,y,t_2)-Y(x,y,t_1)] \nonumber
\eea
with $x_m = \min [z, (\Omega_m-\Omega_{pl})/\Omega_{pl} ]$ and
\bea
Y &=& \frac{u}{4} \big[ \frac{t}{3}(6a-3b+t^2) + \frac{(a-b)^2 \tan^{-1}[t/\sqrt{b}]}{\sqrt{b}}
\big];  \nonumber \\
u  & =& (3/10)y^2; \;\; a = \frac{x-u}{u}; \;\; b = \frac{Q^2_d}{Q^2}.
\label{Y_t_u}
\eea
The dependence on $Y$ is through the restrictions on the domain of integration:
\bea
t_2 &=& 1 + \sqrt{x/u}; \\
t_1 &=& \max \left( \left| 1-\sqrt{x/u} \right|, (z-x)/y \right) < t_2 \,. \nonumber
\label{t1t2}
\eea
For small $z$ and $y$ and $z>>y^2$ we have $F \sim z^3$; if $z<<y^2$, then
$F \sim z^{5/2}$. The entire functional dependence in shown in Fig.~\ref{Fig5} for several values of $Q$ and $Q_d=k_F$.
For large values of $\Omega$ the simplified description, Eqs.~(\ref{one-ph}) and (\ref{one-pl}),
suitable for analytic treatment of long-wave excitations, looses its accuracy because momenta of two excitations
may compensate each other. In this case, Eq.~(\ref{PIRPA}) needs to be replaced with the exact Lindhard function
\cite{Lindhard}.

At small momentum $Q$ this contribution can also be stronger in comparison to the single-plasmon one.
On the one hand, since there is no threshold for emission of particle-hole pairs, the hybrid process can contaminate the single-plasmon measurement. On the other hand, the intensity of this process involves powers of $(\Omega - \omega_{p})$ and thus fades in the vicinity of the single-plasmon threshold at $\Omega_{pl}$. Away from the threshold, this contribution is very broad and can be easily discriminated from sharp peaks.

\section{Conclusions}

We have addressed the problem of indirect RIXS in metals and provided a quantitative framework for understanding the key features of the RIXS signal in Coulomb systems, including two-excitation processes. We have done so based on a field theoretic approach that takes into account long-range Coulombic interactions. As a testbed we have studied the Coulomb gas.  Here we have been able to distinguish single-particle from multi-particle excitations. We have found that the two-plasmon and hybrid signals can be stronger in intensity than the single-excitation spectra at small momentum transfer and that they have different distinctive characteristics.

We have used here an RPA approach, valid for small values of $r_s$ (including $r_s=1$). For larger values of $r_s$ one has to consider diagrams accounting for self-energy, polarization, and vertex function corrections, as well as diagrams that do not factor into the product of $U$ lines. We plan to implement this program within diagrammatic Monte Carlo method in an approximation-free way, see e.g. \cite{EPI2016}.

It is worth mentioning that the same technique can be applied to study the direct RIXS process. The short hole's life time approximation can be removed by computing fully dressed Green's functions obtained in the way described in Ref.~\cite{OC2019}. One will also need to account for the inter-band transitions and distinguish contributions originating from localized holes and mobile $p,d$-electrons.

\section{Acknowledgements}

We are grateful to E. Demler and T. Devereaux for valuable discussions and to K. Gilmore for making extensive comments on the manuscript. AMT and RMK are supported by Office of Basic Energy Sciences, Material Sciences and Engineering Division, U.S. Department of Energy (DOE) under Contract No. DE-SC0012704. IST and NVP are supported by the Simons Collaboration on the Many Electron Problem and the National Science Foundation under the Grant No. PHY-1720465.

\section{Appendix}
\begin{widetext}

 We introduce the notations for our computations below:
 \bea
 G^{(p)}(\omega,k) = [\ri\omega_n - \epsilon(k)]^{-1}, ~~ G^{(s)}(\omega) = [\ri\omega_n + \ri\Gamma\mbox{sign}\omega_n]^{-1}.
 \eea
 
Now there are two vertex diagrams with one dashed line. 
 The simplest diagram with one Coulomb dashed line consists of two parts depending on whether this line is connected to the hole:
 \bea
 && \gamma^{(h)} = 2T\sum_{\omega}\int \frac{\rd^3 k}{(2\pi)^3} G^{(s)}(\omega)G^{(s)}(\omega +\omega_i - \omega_f)G^{(p)}(\omega + \omega_i,k) = \nonumber\\
 &&  2T\sum_n\int  \frac{\rd^3 k}{(2\pi)^3}\frac{1}{\ri(\omega_n +\omega_i) - \epsilon_k}\int \frac{\rd x \rho(x)}{\ri\omega_n -x}\int \frac{\rd y \rho(y)}{\ri(\omega_n+\omega_i-\omega_f) -y}=\nonumber\\
 && 2\ri\int \frac{\rd^3 k}{(2\pi)^3}\theta(\epsilon_k) \int \frac{\rd x \rho(x)}{-\ri\omega_i + \epsilon_k -x}\frac{\rd y \rho(y)}{-\ri\omega_f + \epsilon_k -y} \rightarrow 
 2\int \frac{\nu_p(\epsilon)\rd \epsilon}{(\omega_i + \epsilon_{s} - \epsilon + \ri\Gamma)(\omega_f + \epsilon_{s} - \epsilon + \ri\Gamma)},
 \eea
 where the factor of 2 comes from the summation over spin and 
 \[
 \rho(x) = \frac{1}{\pi}\frac{\Gamma}{\Gamma^2 + (x-\epsilon_s)^2},
 \]
or to the p-electron:
 
 \bea
 && \gamma^{(e)} = 2T\sum_{\omega}\int \frac{\rd^3 k}{(2\pi)^3} G^{(s)}(\omega)G^{(p)}(\omega +\omega_i - \omega_f,k)G^{(p)}(\omega + \omega_f,k-q) = \nonumber\\
&& 2T\sum_n\int \frac{\rd^3 k}{(2\pi)^3} \frac{1}{[\ri(\omega_n +\omega_i) - \epsilon_k][\ri(\omega_n +\omega_f) - \epsilon_{k-q}]}\int \frac{\rd x \rho(x)}{\ri\omega_n -x}=\rightarrow \nonumber\\
 && 2\int  \frac{\rd^3 k}{(2\pi)^3}\frac{1}{\omega_i - \omega_f +\ri\delta - \epsilon_{k-q}+ \epsilon_k}\Big[\frac{\theta(\epsilon_k)}{\omega_f + \ri\Gamma - \epsilon_k + \epsilon_s} - \frac{\theta(\epsilon_{k-q})}{\omega_i + \ri\Gamma - \epsilon_{k-q} + \epsilon_s}\Big] .
\eea
Here the arrows denote the analytic continuation described in the main text.

 Since $\epsilon_k0$ is always positive, we have 
 \bea
  \gamma^{(e)} = 2\int \frac{\rd^3 k}{(2\pi)^3} \frac{1}{(\omega_f + \ri\Gamma - \epsilon_k + \epsilon_s)(\omega_i + \ri\Gamma - \epsilon_{k-q} + \epsilon_s)},
 \eea
 so that the entire vertex becomes
 \bea
 \gamma^{(1)} \equiv \gamma^{(h)} - \gamma^{(e)} = 2\int \frac{\rd^3 k}{(2\pi)^3}\frac{\epsilon_k - \epsilon_{k-q}}{(\omega_f + \ri\Gamma - \epsilon_k + \epsilon_s)(\omega_i + \ri\Gamma - \epsilon_{k-q} + \epsilon_s)(\omega_i + \ri\Gamma - \epsilon_{k} + \epsilon_s)}.
 \eea
 It is universal only at small momenta. Adopting $\epsilon_k = k^2/2m$ we obtain 
 \be
 \gamma^{(1)}(q) =  \frac{q^2}{2m}(2m/\Gamma)^{3/2}\Big[\frac{1}{\pi^2}\int_0^{\infty}\frac{\rd x x^{1/2}}{(\ri -x)^3}\Big].
 \ee
 (the part linear in $q$ cancels in the integration over ${\bf k}$ due to the inversion symmetry). Otherwise the vertex is model dependent.
 
 The second order vertex (with two wavy lines, one with
 $(\omega_1,q_1)$ and another with $(\Omega -\omega_1, Q -q_1)$) is
 (here we have included $\epsilon_s$ in the definition of $\epsilon_k$): 
 \bea
 && \gamma^{(2)} = \gamma^{hh} + \gamma^{ee} - 2\gamma^{he} = 2\int \frac{\rd^3 k}{(2\pi)^3}[\Gamma_1 + \Gamma_2 + \Gamma_3]; \\
 && \Gamma_1 = \frac{1}{[\ri\Gamma - \omega_f - \epsilon_{k- q_f}][\ri\Gamma - (\omega_f +\omega_1) - \epsilon_{k- q_f-q_1}][\ri\Gamma - \omega_i- \epsilon_{k- q_i}]};\nonumber\\
 && \Gamma_2 = -\frac{2\ri\Gamma - (\omega_i +\omega_f) - \epsilon_{k-q_1-q_f}- \epsilon_{k-q_i}}{[\ri\Gamma - \omega_f - \epsilon_{k- q_f-q_1}][\ri\Gamma - (\omega_f +\omega_1) - \epsilon_{k- q_f-q_1}][\ri\Gamma - \omega_i- \epsilon_{k- q_i}][\ri\Gamma +\omega_1 -\omega_i - \epsilon_{k-q_i}]};\nonumber\\
 && \Gamma_3 = \frac{1}{[\ri\Gamma - \omega_f - \epsilon_{k}][\ri\Gamma - (\omega_i -\omega_1) - \epsilon_{k}][\ri\Gamma - \omega_i- \epsilon_{k}]}.\nonumber
 \eea
 When the external momenta are zero it also vanishes. It can be rewritten as 
 \bea
 && \gamma^{(2)} = 2\int \frac{\rd^3 k}{(2\pi)^3}\Big\{\frac{1}{[\ri\Gamma - (\omega_f+\omega_1) - \epsilon_k][\ri\Gamma - \omega_i - \epsilon_{k +Q-q_1}]}\Big(\frac{1}{\ri\Gamma -\omega_f - \epsilon_{k+q_1}}- \frac{1}{\ri\Gamma -\omega_f - \epsilon_{k}}\Big) + \nonumber\\
 && \frac{1}{[\ri\Gamma + (-\omega_i+\omega_1) - \epsilon_k][\ri\Gamma - \omega_i - \epsilon_{k}]}\Big(\frac{1}{\ri\Gamma -\omega_f - \epsilon_{k}}- \frac{1}{\ri\Gamma -\omega_f - \epsilon_{k-Q+q_1}}\Big) \Big\}.
 \eea
 At small momenta, $\gamma^{(2)}(q_1,Q-q_1) \sim [q_1^2 - ({\bf Q}+{\bf q}_1)^2]$. However, this leading contribution cancels when one adds up two diagrams on Fig. 2a and Fig. 2b with interchanged legs:
 \bea
 \gamma^{(2)}(q_1,Q-q_1) + \gamma^{(2)}(Q-q_1,q_1) \sim \int \rd^3 k (\epsilon_k - \epsilon_{k+q_1})(\epsilon_k - \epsilon_{Q-q_1 +k}).
 \eea
 At small momenta, this equals $\sim [{\bf q}_1({\bf Q}- {\bf q}_1)]$.  

\end{widetext}

\end{document}